\documentclass{ws-ijmpd}

\usepackage{graphicx}
\usepackage{epstopdf}

\begin{document}
\markboth{M Govender, K P Reddy and S D Maharaj} {The role of shear in dissipative gravitational collapse}

%
%

\title{The role of shear in dissipative gravitational collapse}

\author{M Govender \footnote{E-mail: govenderm43@ukzn.ac.za}}
\author{K P Reddy\footnote{E-mail: kevinr@dut.ac.za}}
\author{S D Maharaj\footnote{E-mail: maharaj@ukzn.ac.za}}
\address{Astrophysics and Cosmology Research Unit, School of Mathematics, Statistics and Computer Science, \\
University of KwaZulu--Natal, Private Bag X54001, Durban 4000, South Africa}

\maketitle

\begin{history}
\received{Day Month Year}
\revised{Day Month Year}
\end{history}

\begin{abstract}
In this paper we investigate the physics of a radiating star undergoing dissipative collapse in the form of a radial heat flux.
Our treatment clearly demonstrates how the presence of  shear affects the collapse process; we are in a position to contrast the physical features of
the collapsing sphere in the presence of shear with the shear-free case.
By employing a causal heat transport equation of the Maxwell-Cattaneo form we show that the shear leads to an enhancement of the core temperature thus emphasizing that relaxational effects cannot be ignored when the star leaves hydrostatic equilibrium.
\end{abstract}


\section{Introduction}    

The study of relativistic radiating stellar models has been extensive since the presentation of the junction conditions required for the smooth matching of the interior stellar medium to the external atmosphere represented by Vaidya's outgoing solution\cite{santos,vaidya}. Earlier models focussed on shear-free collapse with various assumptions on the internal matter distribution ranging from collapse from an initial static configuration, vanishing of Weyl stresses in the interior and imposing a barotropic equation in the core material. Also the effects of bulk viscosity, anisotropic pressures, cosmological constant and the electromagnetic field have been studied\cite{bon,hs1,t1,herrev1,herrev2}. These `toy' models provide a rich insight into the physics at play during the collapse process, and they have helped investigators understand the dynamical nature of dissipation. Relaxational effects brought about by heat transport through the stellar core were shown to lead to higher core temperatures. In terms of the end-state of collapse, various models have indicated that the formation of
the horizon can be avoided or even delayed in the presence of dissipation.

The inclusion of shear during the collapse process paved the way to study anisotropy in local systems. The first shearing model within this framework was obtained by Naidu {\em et al}\cite{nol}; this model was obtained by ad hoc assumptions on the metric functions. This model was later generalised by Rajah and Maharaj\cite{randm}. Various classes of radiating, shearing models of gravitational collapse were subsequently found, and they exhibited reasonable physical behaviour.  Herrera and
 Santos\cite{herrera10} proposed the so-called Euclidean star model in which the proper radius equals the areal radius throughout the collapse process. Exact solutions for this case lead to physically reasonable models as shown by Govender {\em et al}\cite{govender10}. Chan and co-workers carried out intensive investigations of the role of shear during dissipative collapse\cite{cha1,cha2,cha3,pin}. They were able to show that the inclusion of bulk viscosity diminishes the effective adiabatic index within the stellar core
thereby increasing the instability of the collapsing star. In a recent paper Sharma and Das\cite{dass} considered the role played by anisotropic pressure during the collapse of a radiating sphere executing shear-free motion. They showed that the unequal stresses brought about by pressure anisotropy has a direct impact on the temperature profile of the collapsing star. Recently, Thirukannesh {\em et al}\cite{t2} were able to relax some of the assumptions and were able to provide various classes of shearing, radiating solutions. The simple analytical forms of the solutions allowed them to carry out a physical analysis of a particular model and they were able to show that this model was reasonably well-behaved. Up to this point, all the exact models of radiating stars with shear did not have a shear-free limit, ie., the shear could not be switched off. As a result, it was not possible to highlight the effect of shear directly onto the collapse process. A particular solution of Thirukannesh {\em et al}\cite{t2}(
hereon referred to as {\em TRM}) exhibits the pleasant feature of a shear `switch'.

The purpose of this paper is to highlight the effect of shear on the temperature profiles of the collapsing matter distribution. By utilising the {\em TRM} solution we employ a causal heat transport equation of the Maxwell-Cattaneo form to study the evolution of the temperature when the star leaves hydrostatic equilibrium.

\section{Interior spacetime}

The line element for the interior of the collapsing star is described by the
general spherically symmetric, shearing metric in comoving
coordinates\cite{ggf}
\begin{equation}
ds^2 = -A^2dt^2 + B^2dr^2 + R^2(d\theta^2 +
\sin^2{\theta}d\phi^2) \, ,\label{metric}
\end{equation}
where $A = A(t,r)$, $B = B(t,r)$ and $R =
R(t,r)$. The interior matter content is that of an imperfect fluid given by
\begin{equation} T_{\alpha\beta} = (\mu + P_{\perp})V_\alpha V_\beta + P_{\perp}g_{\alpha\beta} + (P_r - P_{\perp})\chi_\alpha\chi_\beta + q_\alpha V_\beta + q_\beta V_\alpha \, ,\label{2}
\end{equation} where $\mu$ is the energy density, $P_r$ the radial pressure,
$P_{\perp}$ the tangential pressure and $q^{\alpha}$ is the heat
flux vector. The fluid four--velocity ${\bf V}$ is comoving and is
given by
\begin{equation} V^\alpha = \displaystyle\frac{1}{A} \delta^{\alpha}_0 \,.
\label{2'}
\end{equation} The heat flow is in the radial direction and the heat flow vector takes the form
\begin{equation} q^\alpha = (0, q^1, 0, 0)\,, \label{2''}
\end{equation} where $ q^\alpha V_\alpha = 0 $. We further have
\begin{equation}
\chi^{\alpha}\chi_{\alpha} = 1, \hspace{2cm} \chi^{\alpha}V_{\alpha} =0.\end{equation}
The expansion scalar and the fluid four acceleration are given by
\begin{equation}
\Theta = {V^{\alpha}}_{;\alpha}, \hspace{2cm} a_{\alpha} = V_{\alpha;\beta}V^{\beta},
\end{equation}
and the shear tensor is
\begin{equation}
\sigma_{\alpha_\beta} = V_{(\alpha;\beta)} + a_{(\alpha}V_{\beta )} - \frac{1}{3}\Theta (g_{\alpha \beta} + V_\alpha V_\beta). \end{equation}
For the comoving line element (\ref{metric}) the kinematical quantities take the following forms
\begin{eqnarray} \label{kinect}
a_1 &=& \frac{A'}{A}, \label{kinecta} \\
\Theta &=& \frac{1}{A}\left(\frac{\dot B}{B} + 2\frac{\dot R}{R}\right), \label{kinectb}\\
\sigma &=& \frac{1}{A}\left(\frac{\dot A}{A} - \frac{\dot R}{R}\right), \label{kinectc} \end{eqnarray}
 where dots and primes denote differentiation with respect to $t$ and $r$ respectively. The nonzero
components of the Einstein field equations for the line element
(\ref{metric}) and the energy momentum (\ref{2}) are
\begin{eqnarray}
\mu &=& \frac{1}{A^2}\left(2\frac{\dot B}{B} + \frac{\dot R}{R}\right)\frac{\dot R}{R} - \frac{1}{B^2}\left[2\frac{R''}{R} + \left(\frac{R'}{R}\right)^2 - 2\frac{B'}{B}\frac{R'}{R} - \left(\frac{B}{R}\right)^2\right],  \label{14a} \\ \nonumber \\
P_r &=& -\frac{1}{A^2} \left[2\frac{{\ddot R}}{R} - \left(2\frac{\dot A}{A} - \frac{\dot R}{R}\right)\frac{\dot R}{R}\right]  + \frac{1}{B^2}\left(2\frac{A'}{A} + \frac{R'}{R}\right)\frac{R'}{R} - \frac{1}{R^2}, \label{14b}  \\
P_{\perp} &=& -\frac{1}{A^2}\left[\frac{\ddot B}{B} + \frac{{\ddot R}}{R} - \frac{\dot A}{A}\left(\frac{\dot B}{B} + \frac{\dot R}{R}\right) + \frac{\dot B}{B}\frac{\dot R}{R}\right]  \nonumber \\&&+ \frac{1}{B^2}\left[\frac{A''}{A} + \frac{R''}{R} - \frac{A'}{A}\frac{B'}{B} + \left(\frac{A'}{A} - \frac{B'}{B}\right)\frac{R'}{R}\right], \label{14c}\\
q &=& \frac{2}{AB}\left(\frac{\dot{R'}}{R} - \frac{\dot
B}{B}\frac{R'}{R} - \frac{\dot R}{R}\frac{A'}{A}\right),
\label{14d}
\end{eqnarray}
where $q =Bq^1$. This is an underdetermined system of four coupled partial differential
equations in seven unknowns viz., $ A, B, R, \mu, P_r, P_\perp $ and $q$.

\section{Exterior spacetime and junction conditions}

The exterior spacetime is taken to be the Vaidya solution given
by\cite{vaidya}
\begin{equation} \label{v1}
ds^2 = - \left(1 - \frac{2m(v)}{\cal R}\right) dv^2 - 2dvd{\cal R} + {\cal R}^2
\left(d\theta^2 + \sin^2\theta d\phi^2 \right)\,,
\end{equation}  where $m$($v$) represents the Newtonian
mass of the gravitating body as measured by an observer at infinity.
The necessary conditions for the smooth matching of the interior
spacetime (\ref{metric}) to the exterior spacetime (\ref{v1}) have
been extensively investigated. We present the main results that are
necessary for modeling a radiating star. The continuity of the
intrinsic and extrinsic curvature components of the interior and
exterior spacetimes across a time-like boundary give
\begin{eqnarray} \label{j}
m(v)_{\Sigma} &=& \left\{\frac{R}{2}\left[\left(\frac{\dot R}{A}\right)^2 - \left(\frac{R'}{B}\right)^2 + 1\right]\right\}_\Sigma ,\label{j1}\\
(P_r)_\Sigma &=& q_{\Sigma}\label{j2}.\end{eqnarray}
Relation (\ref{j2}) determines the temporal evolution of the collapsing star.

\section{Temporal evolution}

The junction condition $(P_r)_{\Sigma} = q_\Sigma$ yields

\begin{eqnarray}  \label{zz}
{\dot B} = \left(\frac{R}{2AR'}\right)\left[2\frac{\ddot R}{R} +
\left(\frac{\dot R}{R}\right)^2 - 2\frac{\dot A}{A}\frac{\dot R}{R}
+ \frac{A^2}{R^2}\right]B^2 \nonumber \\ + \left[\frac{{\dot R}'}{R'} -
\frac{A'}{A}\frac{\dot R}{R'}\right]B -
\frac{A}{2}\left[\frac{R'}{R} + 2\frac{A'}{A}\right], \end{eqnarray}
 which is of the form
 \begin{equation}\label{zzz}
 {\dot B} = {\cal C}_0(t)B^2 + {\cal C}_1(t)B + {\cal
 C}_2(t).\end{equation}
Thirukannesh {\em et al}\cite{t2} have provided various classes of exact solutions to (\ref{zz}).
Setting
\begin{equation}
\frac{\dot{R}'}{R'}-\frac{A'}{A}\frac{\dot{R}}{R'}=0,
\end{equation}
equation (\ref{zz}) reduces to an inhomogeneous Riccati
equation. Integrating this equation we get
\begin{equation}
A=\alpha(t) \dot{R},
\end{equation}
where $\alpha(t)$ is an integration constant. In this case
(\ref{zz}) becomes
\begin{equation}
\label{a18} \dot{B}=
\left[\frac{\dot{R}(1+\alpha^2)}{2RR'\alpha}-\frac{\dot{\alpha}}{\alpha^2R'}\right]B^2 -\left[\dot{R}'\alpha
+\frac{\dot{R}R'\alpha}{2R}\right].
\end{equation}
This is an inhomogenous Riccati equation which is difficult to
analyse in general.

We take $\alpha$ is a constant and $R$ to be the separable
function
\begin{equation}
\label{a19} R(t,r)=K(r)C(t),
\end{equation}
where  $K(r)$ and $C(t)$ are arbitrary functions of $r$ and $t$
respectively. Then (\ref{a18}) can be written as
\begin{equation}
\label{a20}  \dot{B}= \left[\frac{(1+{\alpha}^2)}{2 \alpha
K'}\frac{\dot{C}}{C^2}\right]
B^2 -\frac{3}{2} \alpha K' \dot{C}.
\end{equation}
For $\alpha = -2$, Thirukkanesh {\em et al}\cite{t2} were able to provide a simple exact solution to (\ref{a20}) of the form
\begin{eqnarray} \label{sol}
A(r,t) &=& -2K{\dot C},\label{sola}\\
B(r,t) &=& \frac{2K'C[3C^4 + f(r)]}{5C^4 - f(r)},\label{solb}\\
R(t,r)&=& K(r)C(t).\label{solc}\end{eqnarray}
where $C, K$ and $f$ are arbitrary functions.
 An inspection of the metric functions $B(r,t)$ and $R(r,t)$ reveals that in the limit $f(r)\rightarrow 0$ leads to vanishing shear. With an insightful choice of $f(r)$ we can investigate the role played by shear directly on the collapse process. This will be taken up in the next section. It is remarkable that we can write down the kinematical and thermodynamical quantities in compact, closed form. The acceleration, collapse rate and shear are given by
\begin{eqnarray} \label{strand}
a &=&\frac{K'(r)}{K(r)},\label{stranda}\\
\Theta &=& -\frac{3f^2(r)+26f(r)C^4(t)-45C^8(t)}{2f^2(r)K(r)C(t)-4f(r)K(r)C^5(t)-30K(r)C^9(t)}, \label{strandb}\\
\sigma &=& -\frac{16f(r)C^3(t)}{K(r)[f(r)-5C^4(t)][f(r)+3C^4(t)]}. \label{strandc}
\end{eqnarray}
The Einstein field equations (\ref{14a})-(\ref{14d}) yield
\begin{eqnarray}
\mu &=& -\frac{l(r,t)}{2K^2(r)C^2(t)([f(r)-5C^4(t)][f(r)+3C^4(t)]^3K'(r)},\\
P_{r} &=& q = -\frac{f^2(r)+30f(r)C^4(t)-15C^8(t)}{2K^2(r)C^2(t)\bigl[f(r)+3C^4(t)\bigr]^2},\\
P_{\perp}&=& \frac{4C^2(t)g(r,t)}{K^2(r)[f(r)-5C^4(t)]^2[f(r)+3C^4(t)]^3K'(r)},
\end{eqnarray} where we have defined
\begin{eqnarray}
g(r,t)&=& K(r)[f(r)-5C^4]^3f'(r)-[f(r)+3C^4(t)][13f^3(r)+45f^2(r)C^4(t) \nonumber \\
&& + 95f(r)C^8(t)-25C^{12}(t)]K'(r), \\
l(r,t)&=& -3f^4(r)K'(r)-52f^3(r)C^4(t)K'(r)+20f(r)C^8(t)[-4K(r)f'(r) \nonumber \\
&& + 7C^4(t)K'(r)]+5C^{12}(t)[40K(r)f'(r)+57C^4(t)K'(r)] \nonumber \\
&& + 2f^2(r)[4K(r)C^4(t)f'(r)-57C^8(t)K'(r)].
\end{eqnarray}

\section{A particular radiating model}
In order to extract the physics of the collapse process we select a particular model in which the radial and temporal evolution are specified as follows:
\begin{eqnarray}
C(t) &=&1+at^2, \\
K(r)&=&1+br^2,\\
f(r)&=&\frac{\gamma}{r^2+c^2},
\end{eqnarray}
where $a$, $b$, $c$ and $\gamma$ are constants. Consequently the metric functions become
\begin{eqnarray} \label{us}
A(r,t)&=& -4a(1+br^2)t, \label{usa}
\\ \nonumber \\
B(r,t)&=&\frac{4br(1+at^2)\biggl[3(1+at^2)^4+\frac{\gamma}{c^2+r^2}\biggr]}{5(1+at^2)^4-\frac{\gamma}{c^2+r^2}}, \label{usb}
\\ \nonumber \\
R(r,t)&=&(a+br^2)(1+at^2). \label{usc}
\end{eqnarray} for the line element (\ref{metric}).
The kinematical quantities (\ref{kinecta})-(\ref{kinectc}) yield
\begin{eqnarray}
a &=&\frac{2br}{1+br^2},\\ \nonumber \\
\Theta &=&-\frac{2+\frac{\bigl[5(1+at^2)^4-\frac{\gamma}{c^2+r^2}\bigr]\bigl[15(1+at^2)^8-\frac{30(1+at^2)^4\gamma}{c^2+r^2}-\frac{{\gamma}^2}{(c^2+r^2)^2}\bigr]}{\bigl[-5(1+at^2)^4+\frac{\gamma}{c^2+r^2}\bigr]^2\bigl[3(1+at^2)^4+\frac{\gamma}{c^2+r^2}\bigr]}}{2(1+br^2)(1+at^2)},
\\ \nonumber \\
\sigma &=&\frac{\bigl[16(c^2+r^2)(1+at^2)^3\gamma\bigr]}{\bigl[(1+br^2)\bigl(5(c^2+r^2)(1+at^2)^4-\gamma\bigr)\bigl(3(c^2+r^2)(1+at^2)^4+\gamma\bigr)\bigr]}.
\end{eqnarray}
The matter variables assume the following form
\begin{eqnarray}
\mu &=&\frac{\Bigl[\frac{8(1+at^2)^4\gamma\bigl[-5(c^2+r^2)(1+at^2)^4+\gamma\bigr]}{b}+\frac{w(r,t)}{\bigl[5(c^2+r^2)(1+at^2)^4-\gamma\bigr]}\Bigr]}{2(1+br^2)^2(1+at^2)^2\bigl[3(c^2+r^2)(1+at^2)^4+\gamma\bigr]^3},
\\
P_r &=& q = -\frac{-15(1+at^2)^8+\frac{30(1+at^2)^4}{c^2+r^2}+\frac{{\gamma}^2}{(c^2+r^2)^2}}{2(1+br^2)^2(1+at^2)^2\bigl[3(1+at^2)^4+\frac{\gamma}{c^2+r^2}\bigr]^2},
\\
P_{\perp}&=&\frac{z(r,t)}{\bigl[b(1+br^2)^2(-5r^2(1+at^2)^4+\gamma)^2(3r^2(1+at^2)^4+\gamma)^3\bigr]},
\end{eqnarray}
where the function $w(r,t)$ is defined as
\begin{eqnarray}
w(r,t)&=&\Bigl[285c^8(1+at^2)^{16}+285r^8(1+at^2)^{16}-60r^2(1+at^2)^{12}\gamma \nonumber \\
&& -34r^4(1+at^2)^8{\gamma}^2-60r^2(1+at^2)^4{\gamma}^3-3{\gamma}^4+20c^6(1+at^2)^{12} \nonumber \\
&& \times(57r^2(1+at^2)^4+7{\gamma})+2c^4(1+at^2)^8\bigl[855r^4(1+at^2)^8 \nonumber \\
&& +110r^2(1+at^2)^4\gamma-57{\gamma}^2\bigr]+4c^2(1+at^2)^4\times\bigl[285r^6(1+at^2)^{12} \nonumber \\
&& +5r^4(1+at^2)^8\gamma-37r^2(1+at^2)^4{\gamma}^2-13{\gamma}^3\bigr]\Bigr],
\end{eqnarray}
and $z(r,t)$ is defined as
\begin{eqnarray}
z(r,t)&=&\Bigl[4(1+at^2)^2\Bigl[(5r^2(1+at^2)^4-\gamma)^3\gamma+br^2\bigl[75r^8(1+at^2)^{16} \nonumber \\
&& -135r^6(1+at^2)^{12}\gamma-305r^4(1+at^2)^8{\gamma}^2-69r^2(1+at)^4{\gamma}^3 \nonumber \\
&& -14{\gamma}^4\bigr]\Bigr]\Bigr].
\end{eqnarray}
The thermodynamical quantities at the centre of the star are given by
\begin{eqnarray}
q_0&=&-\frac{-15(1+at^2)^8+\frac{30(1+at^2)^4\gamma}{c^2}+\frac{{\gamma}^2}{c^4}}{2(1+at^2)^2\bigl[3(1+at^2)^4+\frac{\gamma}{c^2}\bigr]^2},
\\ \nonumber \\
(P_{r})_0&=&\frac{-5+\frac{3\bigl[-5(1+at^2)^4+\frac{\gamma}{c^2}\bigr]^2}{\bigl[3(1+at^2)^4+\frac{\gamma}{c^2}\bigr]^2}}{4(1+at^2)^2},
\\ \nonumber \\
\mu_0&=&\frac{\Bigl[\frac{8(1+at^2)^4\bigl(-5c^2(1+at^2)^4+\gamma\bigr)}{b}+\frac{w(0,t)}{\bigl(5c^2(1+at^2)^4-\gamma\bigr)}\Bigr]}{2(1+at^2)^2\bigl[3c^2(1+at^2)^4+\gamma\bigr]^3},
\\ \nonumber \\
(P_{\perp})_0&=&-\frac{4(1+at^2)^2}{b\gamma},
\end{eqnarray}
where $w(0,t)$ is given by
\begin{eqnarray}
w(0,t)&=&285c^8(1+at^2)^{16}+140c^6(1+at^2)^{12}\gamma-114c^4(1+at^2)^8{\gamma}^2 \nonumber \\
&& -52c^2(1+at^2)^4{\gamma}^3-3{\gamma}^4,
\end{eqnarray}
and subscripts  denote the quantities evaluated at $r = 0$.

\section{Causal thermodynamics}

Causal heat flow in dissipative collapse has received widespread attention since the pioneering work by Herrera and co-workers\cite{herrev4,herrev5,herrev6,herrev7} on radiating stars. Many of the early stellar models described shear-free, dissipative collapse in the form of a radial heat flow in the free-streaming or diffusion approximation. Results show that relaxational effects are prominent during the latter stages of collapse. In particular, it was shown that relaxational effects lead to higher temperatures within the stellar interior with cooling being enhanced closer to the surface of the collapsing body. The effect of shear and pressure anisotropy were recently highlighted within the Eckart framework. It is well known that the Eckart formalism of dissipative processes has several shortcomings as pointed out in numerous studies. The noncausal nature of the theory as well as its prediction of unstable equilibrium states has warranted the use of casual transport equations of Maxwell-Cattaneo form. The study of relaxational effects during radiating, shearing collapse using a causal heat transport equation was first carried out by Di Prisco {\em et al}\cite{herrev3}. Their results show that the luminosity profiles obtained for nonvanishing relaxation times are different from those obtained when the relaxation times are ignored.
The relativistic version of the Cattaneo equation for the heat transport assumes the following form
\begin{eqnarray}
\tau_1 {h_\alpha}^\beta {\dot{q}}_\beta + q_\alpha &=& - \kappa (D_\alpha T +
T {\dot{u}}_\alpha), \label{r2.26}\\
\tau_2 {h_\alpha}^\mu {h_\beta}^\nu {\dot{\pi}}_{\mu \nu} &=& -2 \eta \sigma_{\alpha \beta}, \label{r2.27}
\end{eqnarray} where the relaxation times $\tau_A (\rho, n)$ are given by
\begin{equation} \label{r2.24}
\tau_1 = \kappa T \beta_1,\,\,\,\,\, \tau_2 = 2 \eta \beta_2,
\end{equation} for the heat flux and shear viscosity respectively.
The evolution terms, with the relaxation time coefficients $\tau_A$,
are needed for causality, as well as for modelling high-frequency
or transient phenomena, where `fast' variables and relaxation
effects are important\cite{herrev8,herrev9,herrev10,herrev11,maartens96}.  It has been shown for spherically symmetric
stars with radial heat flow, the temperature gradient which
appears as a result of the perturbation, and hence the luminosity,
are highly dependent on the product of the relaxation time by the
period of the oscillation of the star. For the line element (\ref{metric}) the causal heat transport
equation (\ref{r2.26}) becomes \begin{equation} \label{ca1}
\tau(qB)_{,t} + A(qB) = -\kappa \frac{(AT)_{,r}}{B},
\end{equation} which governs the behavior of the temperature.
Setting $\tau = 0$ in (\ref{ca1}) we obtain the familiar Fourier
heat transport equation \begin{equation} \label{ca2} A(qB) =
-\kappa \frac{(AT)_{,r}}{B}, \end{equation} which predicts
reasonable temperatures when the fluid is close to
quasi--stationary equilibrium.

The thermodynamic coefficients for radiative transfer were well motivated by Govender\cite{megan} in which it was assumed that heat dissipation
occurred via thermally generated neutrinos moving radially outwards from the stellar core. The thermal
conductivity takes the form \begin{equation} \kappa =\chi
T^3{\tau}_{\mathrm c} \label{a28}\,,\end{equation} where $\chi$ ($\geq0$) is a constant and ${\tau}_{\mathrm c}$ is the mean
collision time between the massless and massive particles. Based
on this treatment we assume the power--law behavior
\begin{equation} \label{a29} \tau_{\mathrm c}
=\left({\alpha\over\chi}\right) T^{-\omega} \,,\end{equation}
where $\alpha$ ($\geq 0$) and $\omega$ ($\geq 0$) are constants.
As was shown by Martinez\cite{mart} the case $\omega={3\over2}$ corresponds to thermally generated
neutrinos in neutron stars. The mean collision time decreases with
growing temperature, as expected. For the special case
$\omega=0$, the mean collision time is constant. We assume that relaxation time is proportional to the
collision time: \begin{equation} \tau =\left({\beta \chi \over
\alpha}\right) \tau_{\mathrm c} \label{a30}\,,\end{equation} where
$\tau$ ($\geq 0$) is a constant.

Using the above definitions for $\tau$ and $\kappa$, (\ref{ca1})
takes the form \begin{equation} \beta (qB)_{,t} T^{-\omega} + A (q
B) = - \alpha \frac{T^{3-\omega} (AT)_{,r}}{B} \label{temp1}
\,.\end{equation}
In the case of constant mean collision time, {\it ie.} $\omega=0$,
the causal transport equation (\ref{temp1}) is simply integrated to
yield\cite{keshlin}
\begin{equation} (AT)^4 = - \frac{4}{\alpha} \left[\beta\int A^3 B
(qB)_{,t}{\mathrm d} r + \int A^4 q B^2 {\mathrm d} r\right] +
F(t) \label{caus0} \end{equation} where $F(t)$ is an integration function. In order to investigate the
relaxational effects due to shear we utilize (\ref{r2.27}) as a
definition for the relaxation time for the shear stress. For the
metric (\ref{metric}) the shear transport equation (\ref{r2.27}) reduces
to
\begin{equation}
{\tau}_1 = \frac{-P}{\dot{P} + \frac{8}{15}r_0\sigma
T^4},\end{equation} where we have used the coefficient of shear
viscosity for a radiative fluid\cite{maartens96}
\begin{equation}\label{eta}
\eta = \frac{4}{15}r_0T^4\tau_1,\end{equation} where $P=
\frac{1}{3}\left(P_\perp - P_r\right)$ and $r_0$ is the
radiation constant for photons. Utilising (\ref{usa})-(\ref{usc}) in (\ref{caus0}) we obtain the causal temperature profile for the case of constant collision time:
\begin{eqnarray}
T^4(r,t)&=& \frac{1}{16K^4(r)}\Biggl(-1/(\psi C^2(t))64\Bigl(\beta\int \Bigl[K(r)(f^3(r)-15f^2(r)C^4(t)\nonumber \\
&& +315f(r)C^8(t)-45C^{12}(t)\Bigr]K'(r)/\Bigl[[f(r)-5C^4(t)][f(r)+3C^4(t)]^2\Bigr]dr \nonumber \\
&& +\Bigl(\int\frac{K^2(r)[f^2(r)+30f(r)C^4(t)-15C^8(t)]K'(r)}{[f(r)-5C^4(t)][f(r)+3C^4(t)]}dr\Bigr)C(t)\Bigr) \nonumber \\
&& +\frac{F(t)}{[C'(t)]^4}\Biggr),
\end{eqnarray}
where $\psi$ is a constant. The function $F(t)$ is evaluated by invoking the boundary condition
\begin{equation} \label{f10}
\left({T^4}\right)_{\Sigma} =
\left(\frac{1}{r^2B^2}\right)_{\Sigma}\left({L_{\infty}\over 4\pi\delta}
\right),
\end{equation}
where $L_{\infty}$ is the total luminosity at
infinity and $\delta$ ($>0$) is a constant.

Fig. 1 clearly shows that the inclusion of shear leads to a higher core temperature. As with previous investigations, Fig. 2 shows that relaxational effects predict a higher temperature than its noncausal counterpart. The interesting feature in our investigation is the enhancement of the relaxational effects brought about by the inclusion of shear. We can understand the increased temperature within the stellar core on the basis that the shear is responsible for the internal friction between the layers of the stellar fluid. This interaction between the layers at each interior point of the collapsing body leads to the generation of heat within the core. Furthermore, we expect the generation of heat due to friction to be more pronounced closer to the centre of the star as the density and pressure here are maximum. Fig. 3 illustrates the trend in the luminosity at infinity as a function of time. For our particular model, the luminosity is independent of time in the shear-free limit. This implies
that in the shear-free model the horizon is never encountered. In this case the rate of energy dissipation is equal to the collapse rate of the core. Similar models were studied by Banerjee {\em et al} \cite{bcd}, Naidu and Govender\cite{nol1} and Maharaj {\em et al}\cite{new}. Fig. 3 shows that the luminosity increases as the shear increases. We note that in our model there is no direct link between the luminosity and the product $\sigma\eta$ where $\eta$ is the shear viscosity coefficient. The trend in the luminosity profile in relation to the shear can be understood in terms of the deviations from a shear-free profile. It has been shown that the scalar function\cite{ospino1}
\begin{equation} \label{ytf}
Y_{TF} = 16\pi\eta\sigma + \frac{4\pi}{R^3}\int_0^R{R^3\left(D_R{\mu} - 3{q}\frac{U}{RE}\right)dR}\end{equation}
 where $D_R = \frac{1}{R^{\prime}}\frac{\partial}{\partial{r}}$ is the proper radial derivative, $D_T = \frac{1}{A}\frac{\partial}{\partial{t}}$ is the proper time derivative, $U = D_TR < 0$ and
 \[
 E = \left(1 + U^2 - \frac{2m}{R}\right)^{1/2}\]
 is the mass function measures the stability of the shear-free condition. It is quite clear that even in the absence of shear viscosity (if we ignore the internal friction between the layers of the collapsing fluid) there can be an increase in the absolute value of the shear scalar due to the presence of density inhomogeneity and heat flux.
 Fig. 4 shows the behaviour of the relaxation time as a function of the radial coordinate. The relaxation time for
the shear stress exhibits substantially different behaviour when
the fluid is close to hydrostatic equilibrium as opposed to
late-time collapse. In particular, our results confirm earlier findings by Naidu {\em et al}\cite{nol} in which they showed that \begin
{equation}\frac{(\tau_{1})_{early}} {(\tau_{1})_{late}} \approx
100.\end{equation} In their model, the acceleration vanishes and the particle trajectories are geodesics. Our work has clearly demonstrated the impact of shear
on observable properties of a radiating, collapsing star. We were able to achieve this because our model allows us to switch off the shear and compare the physics in both the shearing and shear-free limits. In addition, figure 5 shows that the collapse rate at the boundary of the star is higher in the shearing model as compared to its shear-free counterpart. The opposite trend is observed at the centre of the collapsing star, ie., the collapse rate in the presence of shear is lower than the collapse rate in the shear-free case as evidenced in figure 6. It is well known that shear-free collapse proceeds at the slowest possible rate\cite{bon}. Figures 5 and 6 imply that shear is directly responsible for the evolution of energy density inhomogeneities as the fluid collapses. From (\ref{ytf}) we see that energy density inhomogeneity is responsible for deviations from an initially shear-free profile. These deviations can be enhanced in the presence of heat dissipation within the stellar core. The role of energy density inhomogeneity during dissipative collapse was highlighted by Herrera \cite{inhomog} in which it was shown that the factors responsible for the generation of energy density inhomogeneities depends on the nature of the fluid distribution. Our results confirm the findings of Herrera in that both shear and dissipative fluxes (in our case heat flow) lead to the generation of energy density inhomogeneities. It has been shown that sufficiently dominant shearing effects in the neighbourhood of the singularity can delay the formation of the apparent horizon\cite{jdm}. This dispersive effect of shear can lead to the singularity being exposed to an external observer.

\section{Conclusion}

Utilising a solution recently discovered by Thirukannesh {\em et al}\cite{t2} which describes a radiating sphere undergoing dissipative collapse we were able to clearly demonstrate the effect of shear on observable quantities such as the temperature and luminosity profiles. The simple nature of this particular exact solution allows the shear to be switched off which makes it possible (for the first time) to compare the dynamics of the collapse process both in the shearing and shear-free limits. We are able to show that the shear leads to an enhancement of the temperature at each interior point of the collapsing body. For the first time, we are able to show that relaxational effects are pronounced in the presence of shear. These effects are strengthened during late stages of gravitational collapse.
\section*{Acknowledgments}
The authors are grateful to the referee for useful and constructive comments which helped clarify the main results of the paper.

\begin{figure}
\centering
\includegraphics[scale=0.6]{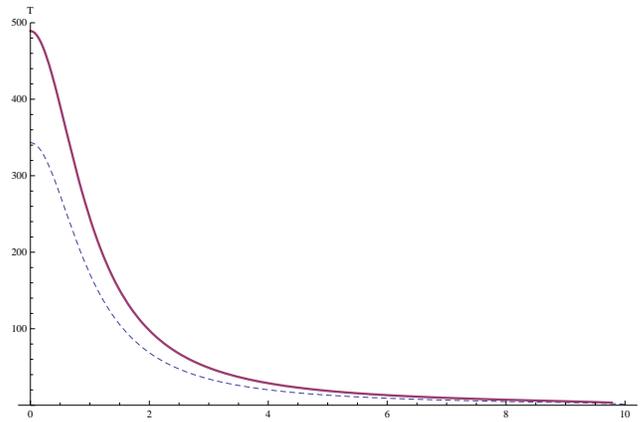}\caption{Noncausal temperature vs radial coordinate} \label{fig1}
\end{figure}

\begin{figure}
\centering
\includegraphics[scale=0.6]{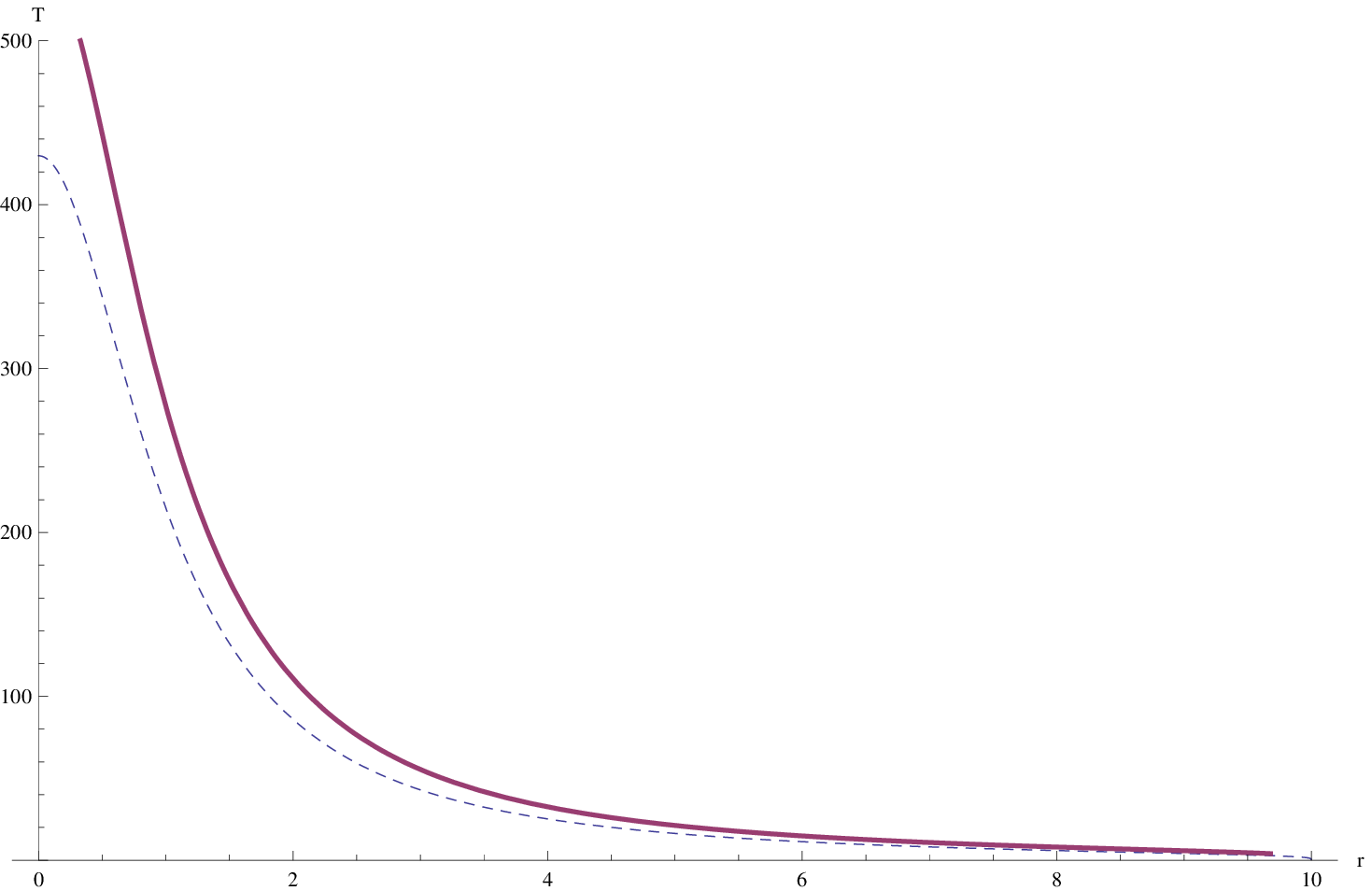}\caption{Causal temperature vs radial coordinate} \label{fig2}
\end{figure}

\begin{figure}
\centering
\includegraphics[scale=0.6]{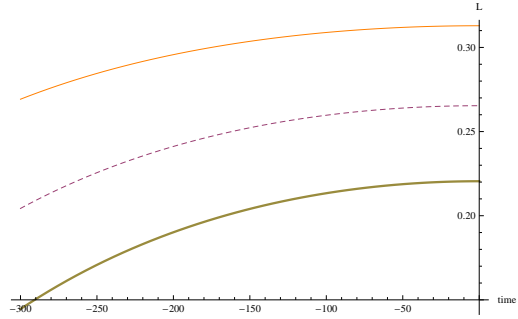}\caption{Luminosity at infinity vs time} \label{fig3}
\end{figure}

\begin{figure}
\centering
\includegraphics[scale=0.6]{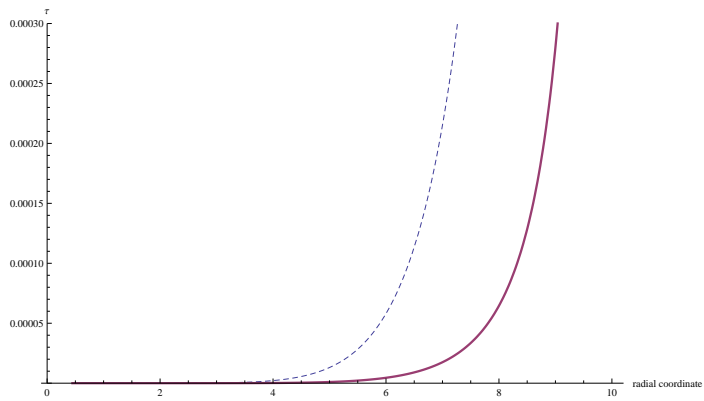}\caption{Relaxation time as a function of the radial coordinate (dashed line - close to equilibrium, solid line - far from equilibrium)} \label{fig4}
\end{figure}

\begin{figure}
\centering
\includegraphics[scale=0.6]{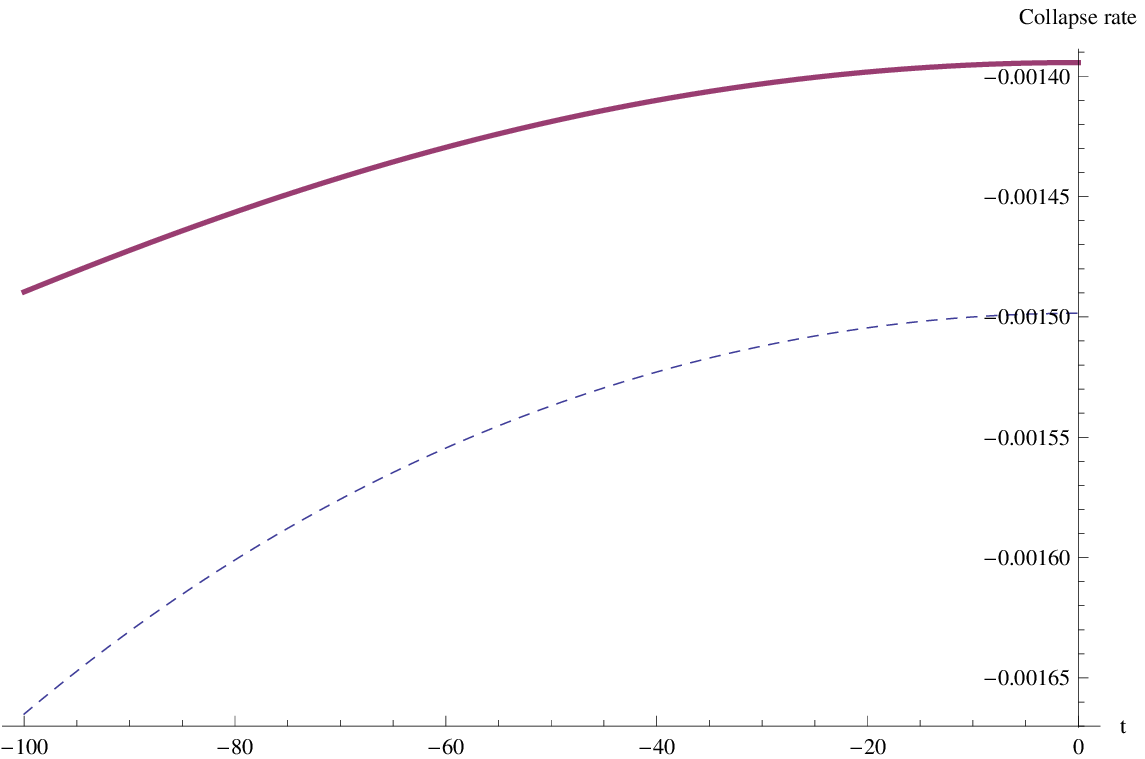}\caption{Collapse rate (at the surface) as a function of time (solid line - shearing collapse rate, dashed line - shear-free collapse rate)} \label{fig5}
\end{figure}

\begin{figure}
\centering
\includegraphics[scale=0.6]{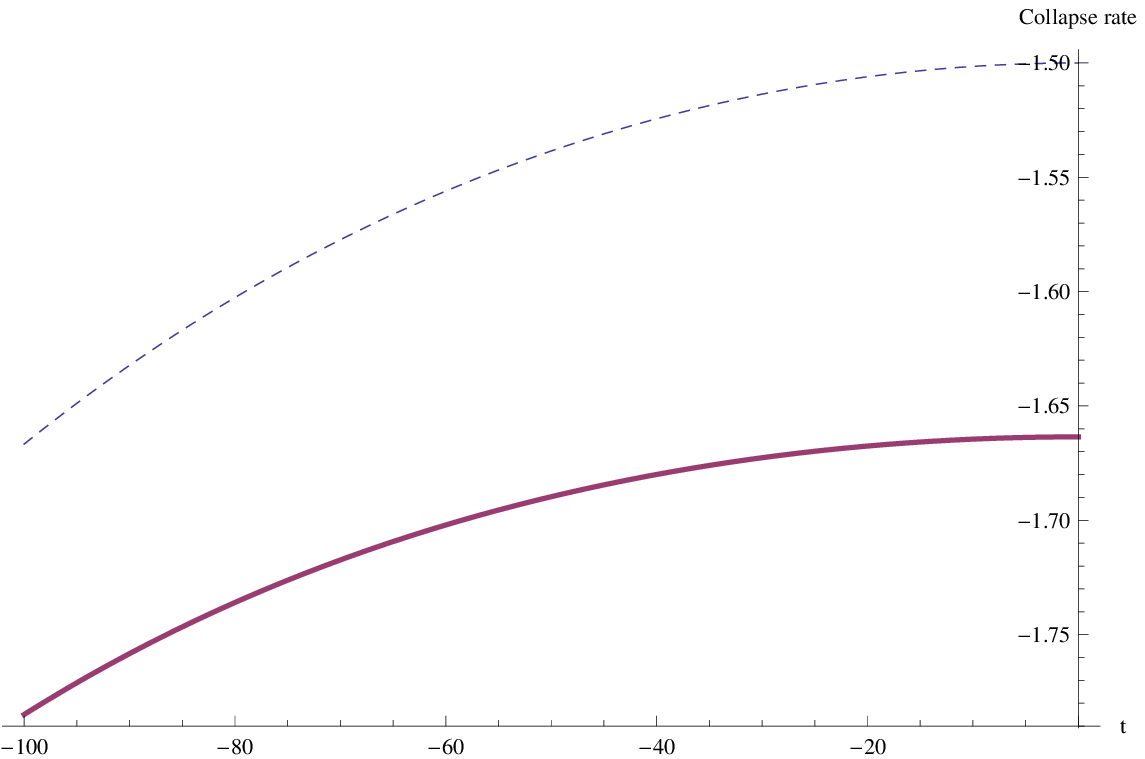}\caption{Collapse rate (at the centre) as a function of time (solid line - shearing collapse rate, dashed line - shear-free collapse rate)} \label{fig6}
\end{figure}
\end{document}